\documentstyle[preprint,prl,aps]{revtex}
\begin{document}
\draft
\title{Phase transitions and the internal noise structure of nonlinear Schr\"odinger equation solitons}
\author{M. J. Werner and S. R. Friberg}
\address{NTT Basic Research Laboratories, 3-1 Morinosato-Wakamiya, Atsugi-shi,Kanagawa-ken,
243-01,Japan.}
\date{\today}
\maketitle
\begin{abstract}
We predict phase-transitions in the quantum noise characteristics of systems
described by the quantum nonlinear Schr\"odinger equation, showing them to be related 
to the solitonic field transition at half the fundamental soliton amplitude.
These phase-transitions are robust with respect to Raman
noise and scattering losses. 
We also describe the rich internal quantum noise
structure of the solitonic fields in the vicinity of the phase-transition.
For optical coherent quantum solitons, this leads to the prediction that eliminating
the peak side-band noise due to the electronic nonlinearity of silica fiber by spectral filtering
leads to the optimal photon-number noise reduction of a fundamental soliton.
\end{abstract}
\narrowtext

An initially localized wavepacket oscillates or "breathes" as it evolves
into a soliton in a system described by a nonlinear Schr\"odinger equation.
For the special case of integral multiples of the fundamental soliton amplitude,
this oscillation is periodic. Otherwise, the wavepacket asymptotically evolves into an
integral order "higher-order" soliton or if
the initial wavepacket energy is near that of a fundamental soliton, the oscillatory breathing
motion decays.
The underlying physics is that the spectrum broadens 
due to the self-phase modulation
and the dispersion acts as a feedback mechanism to 
redirect energy to the pulse center.

A characteristic signature of soliton formation for pulses more energetic
than the fundamental soliton (that is, for $N>1$) is the bifurcation 
and rejoining that occurs in the wavepacket spectral domain.
Breathing oscillations occur for $N<1$ 
but no distinct bifurcation in the intensity spectrum occurs.
(Note, we take $N=1$ as the amplitude of a fundamental soliton.)
When bright solitons do not exist (e.g., for normal dispersion), there is 
no feedback mechanism and the energy is transported away from the pulse center. 
Similarly, for $N<0.5$ in the anomalous dispersion regime, no soliton emerges in the
asymptotic field.
An important property of the soliton system is that for $N>0.5$
the asymptotic field contains one or more fundamental solitons~\cite{ist}. 
The nonlinear Schr\"odinger equation therefore
predicts a phase-transition 
in the {\em asymptotic} field at $N=0.5$ from $0\rightarrow 1$ solitons.
The soliton physics which leads to this behavior has important consequences 
which have not been previously explored. For example, below the phase-transition, quantum effects
are small, whereas above the transition, they play an important role in structuring
the noise properties of optical pulses.

Recent quantum noise experiments with optical solitons in silica fiber used coherent 
mode-locked laser pulses as their source~\cite{nature}.
Though the pulses contained macroscopically 
large numbers of photons, clear and unambiguous
quantum effects were found in the noise measurements. These arose as the initial solitonic fields
contained shot-noise fluctuations that were
incorporated into the soliton and the accompanying dispersive radiation.
In the initial propagation regime, where quantum soliton noise experiments to now have been
carried out, the interference between the emergent soliton 
and dispersive radiation figures significantly in the difference
between classical and quantum soliton field theories.
For example, quantum mechanically, oscillations are necessarily present 
when the initial state is coherent even when
classically they are absent, as when the initial pulse is an unperturbed fundamental soliton.
So far, only one quantum noise experiment~\cite{friberg} 
has reported direct experimental consequences of these oscillation effects --
the oscillations in the photon-number noise of the spectrally filtered solitons
as the soliton energy is increased. 

This Letter briefly describes some of the new physics that emerges from consideration
of the quantum-mechanical effects of solitonic oscillatory behavior.
Perhaps most significant is the phase-transition in the spectral intensity noise of the solitonic
field accompanied by a bifurcation in its noise spectrum as the energy approaches that of a fundamental soliton.
The signature of this transition is predicted to occur 
in the experimentally accessible regime of a few soliton periods, 
even at room-temperature where the Raman noise in silica fiber is important 
for typical picosecond and sub-picosecond pulse durations.
We also detail aspects of the internal quantum noise structure of a propagating, 
oscillating soliton.

The nonlinear Schr\"odinger equation has been used in various forms 
for the study of Bose-Einstein condensation~\cite{bradley95}, 
Bose superfluids~\cite{josserand95}, and propagating 
coherent quantum solitons~\cite{friberg,firstprl,lai,kaertner,werner} where
the nonlinear Schr\"odinger equation governs the dynamics of the photon flux amplitude. 
Its quantum-mechanical description utilizes a normal-order
representation of the quantum fields 
so that losses introduced by Rayleigh scattering and spectral
filtering do not explicitly introduce extra noise into the equations at zero temperature.
In the normal-ordered positive-P representation, the evolution takes on the familiar form
of a zero-temperature ($T=0$ K) Ito stochastic damped 
quantum nonlinear Schr\"odinger equation (QNLSE)~\cite{firstprl} 
in a co-moving frame
\begin{equation}
{{\partial \phi}\over{\partial\zeta}} = \left[-\gamma
-\frac{i}{2}\left(1\pm {{\partial^2}\over{\partial\tau^2}}\right)+
i\phi^\dagger\phi+\sqrt{i}\Gamma(\zeta,\tau)\right]\phi \ ,
\end{equation}
where $\Gamma$ is a real Gaussian noise with zero mean and a correlation
$\langle \Gamma (\zeta, \tau) \Gamma (\zeta ', \tau ')\rangle
= \delta ( \zeta - \zeta ') \delta (\tau - \tau ') / \bar{n}$,
$\bar{n}$ is a dimensionless photon number scale, and the length
and time variables ($\zeta, \tau$) are the scaled coordinates in a
reference frame that moves with the propagating field at the group
velocity of the center frequency of the soliton.  For this
equation and the corresponding Hermitian conjugate equation for
$\phi^\dagger$, the length scales as $t_0^2 / |k''|$, with $t_0$
the pulse width and $k''$ the fiber's group velocity dispersion. 
Quantum field propagation is performed numerically~\cite{wernerd} using the Raman
modified quantum nonlinear Schr\"odinger equation in the positive-P
representation~\cite{carterd91,gawbs}.  
The output field photon number (scaled to the photon-number scale $\bar n$ which for results
presented later is $10^8$) is, after
spectral filtering
\begin{equation} \langle \hat n \rangle  
= \int d \omega f^*(-\omega) f ( \omega ) \langle \hat I(\omega) \rangle 
\end{equation}
where $\hat I(\omega) = \hat \phi^\dagger (-\omega) \hat \phi(\omega)$
and $f(\omega)$ is the spectral filter function~\cite{ramanfilter}, specifically an ideal pass-band filter
in the following calculations.

If one were to measure the spectral intensity noise of the output solitonic field, it would reveal a bifurcation
or splitting of the noise spectrum as the input energy is increased. 
We illustrate this effect in Fig.~1 by plotting the intensity spectrum variance for several input
energies after 4 soliton propagation periods. 
The shot-noise level in all figures is at the zero level due to the normal-ordered representation, and 
an arbitrary scale has been used on the vertical axis to emphasize
the spectral structure. The horizontal axis is scaled to the soliton pulse width
$t_0$ so that the noise structure is independent of the soliton pulse duration
within the limit of applicability of the unmodified QNLSE (for example, no Raman effect).
In addition to the classical bifurcation in the spectrum for $N>1$, the soliton
system displays a bifurcation in the quantum noise spectrum for $N<1$.
The transition from a single noise peak at $N=0.7$ to 
sub-shot noise fluctuations
and a split noise spectrum for $N=0.9$ can be revealed by photon-number noise measurements after spectral filtering. 
At fundamental soliton energies ($N=1$), the noise spectrum contains excess noise in the side peaks similar to $N=0.9$,
shot-noise fluctuations at zero frequency offset and sub-shot noise fluctuations in between. 

Optimizing the quantum noise reduction at each
energy by varying both the spectral filter bandwidth and propagation distance over the interval 
$\xi \in (0,\xi_{max}=4]$ ($\xi$ is given in units of soliton periods) reveals a phase transition 
in the quantum noise correlations of the solitonic field. 
This transition depends on the maximum propagation distance as shown in Fig.~2 where it is depicted for 
$\xi_{max}=4,8$.
Optimization gives the maximal quantum correlation expected for the solitonic field
and therefore reflects the internal quantum noise structure.
The filter cut-off frequency varies from $1/10t_0$ within the transition region to $3/8t_0$ below the transition
for $\xi_{max}=4$.
Note that for $\xi_{max}=8$, the transition occurs at slightly lower energies
as expected for longer propagation distances. 
In contrast, increasing the fiber losses moves the phase transition to higher energies.
Figure 2 also shows the transition to be robust with respect to Raman
noise and the intrinsic scattering losses of a
silica fiber (assuming $T=300K,\gamma=0.0236$, and $\xi_{max}=4$). 
In the normal dispersion regime (where no feedback or breathing oscillations occur), wavepackets 
with the same parameters only show a gradual linear decrease in noise as the  energy increases in the vicinity
of the soliton phase-transition. Consequently, in the normal dispersion regime the optimal 
for all energies shown in Fig. 2 is at $\xi=4$. 
The optimized quantum-noise reduction in the normal dispersion case always displays
sub-shot noise photon-number fluctuations,
whereas the soliton remains at the shot-noise
level below the transition. Our experimental demonstration of  soliton squeezing\cite{friberg}, showed that
lower energy pulses exhibited excess noise after spectral filtering. However, the results presented here
show that excess noise for sub-fundamental soliton energies is not optimal.
Summarizing, the classical $0\rightarrow 1$ soliton transition at $N=0.5$ in the {\em asymptotic} field 
manifests itself in the quantum noise for propagation distances of a few soliton periods. 
This represents a new feature for systems described by the nonlinear Schr\"odinger equation.

The quantum noise spectral structure shown in Fig.~\ref{fig3} that develops with the propagation 
of fundamental solitons  illustrates the striking difference between 
classical and quantum descriptions, even for photon numbers as large as $10^8$.
Classical theory says that fundamental solitons only undergo a global phase shift as they propagate. 
However, Fig.~\ref{fig3} shows that their intensity noise spectrum undergoes
considerable change even after only a few soliton periods, assuming a coherent 
sech$(t/t_0)$ input pulse\cite{heuristic}.
The internal noise structure of the propagating solitonic field, which can be probed by measuring the 
photon number fluctuations of spectrally filtered pulses, provides new insights into soliton dynamics.
One important unexpected new feature is the production of excess noise in the spectral side-bands that
peaks at side-band frequencies of $0.125/t_0$ Hz. This excess  noise,  shown in red in Fig.~\ref{fig3},
can easily be removed by a spectral filter. 
Accompanying this excess noise are regions of sub-shot noise fluctuations, shown as dark blue in Fig.~\ref{fig3},
which are correlated with the excess noise. 

The photon-number squeezing and the phase-transition is perhaps most easily explained by visualizing
the solitons internal quantum noise structure and considering the effect of a spectral filter.
Experimentally, the quantum correlations between different components of the noise  
can be determined by varying the spectral filter
bandwidth and center frequency while measuring the changing noise variance of the transmitted energy.
Figure \ref{fig4} shows the  changes in the photon number 
statistics of filtered fundamental solitons resulting from changing the filter bandwidth and propagation distance.  
For a wide filter bandwidth (i.e., much larger than $0.125/t_0$ Hz), the overall noise is 
near the shot-noise level.
Decreasing the filter bandwidth reduces the noise level to a minimum 
and then increases it again as the filter cut-off removes an increasing portion of the pulse energy. 
In  Fig.~\ref{fig4}
the temperature of the phonon reservoir is $300K$ and $t_0=1$ ps. 
Over the first four soliton periods, 
there is a clear optimum for the choice of filter cut-off frequency:
$125\pm 12.5$ GHz at a propagation 
distance of $3\pm 0.2$ soliton periods. 
This is exactly the same frequency as Fig.~\ref{fig3} for the 
peak of the spectral intensity noise variance due to the electronic nonlinearity.
As Raman effects are included, there is an asymmetry in the spectral intensity noise spectrum
and an increase in the spectral intensity fluctuations with increasing temperature.
Despite this effect, Fig.~\ref{fig4} shows significant photon-number squeezing up to 4.8dB below shot-noise,
even at room-temperature.

The noise spectrum for a pulse with slightly more energy 
than a fundamental soliton has a different noise structure than that of a fundamental soliton.
This can lead to dramatic changes in the measured noise levels after spectral filtering.
Figure \ref{fig5} shows the noise spectrum when $N=1.1$, clearly showing excess noise near the center frequencies,
in constrast to the  $N=1$ case (Fig.~\ref{fig3}) where the fluctuations were at or below shot-noise level\cite{numerical}. 
This leads to large excess noise levels for strongly filtered pulses, which was observed experimentally~\cite{friberg}. 
Despite the small increase in input energy (21\%), the resulting quantum noise properties
are very different and this illustrates the 
importance of understanding the complexity of the quantum noise
structure of these solitonic fields.

Summarizing, we  predict a phase-transition in systems governed by  nonlinear
Schr\"odinger equations detectable by measuring the particle number fluctuations
in a selected subspace of the total Hamiltonian. The choice of a sub-space corresponding to the spectral 
filtering of modes farthest away from the slowly-varying envelope carrier frequency leads to a transition 
 in the filtered pulse photon-number fluctuations which
corresponding to the phase-transition in the quantum noise of the propagating solitonic field.
(This squeezing transition in the co-moving frame, incidentally, 
has similarities to evaporative cooling in
Bose-Einstein condensation.)
We further showed that
this transition is related to the classical  transition to soliton propagation
 that occurs at $N=0.5$
and to the  bifurcation in the intensity noise spectrum, and that it can be seen at easily accessible 
propagation distances.
The internal quantum noise structure of the soliton field was shown to differ qualitatively
depending on whether soliton energies were above or below that of the phase-transition and 
also a fundamental soliton. The  optimal filter bandwidth for reducing photon-number fluctuations for fundamental
solitons  was found to correspond to removal of the peak in the electronic nonlinearity induced
intensity noise spectral side-bands.

We have restricted our attention to systems near the $0\rightarrow 1$ soliton phase-transition
where Raman effects in silica fiber do not play a dominant role for  picosecond pulses.
However, higher-order nonlinearities and dispersion may effect the quantum noise propagation of 
shorter pulses and will be the subject of further studies.

\begin{figure}
\caption{Scaled intensity spectrum variance for $N=0.7$,$0.9$,$1.0$ at $\zeta=2\pi$ using 
$\phi(0,\tau)=N\mbox{sech}(\tau)$. 
\label{fig1}}
\end{figure}
\begin{figure}
\caption{The quantum noise phase transition as the input energy is increased
depicted using the photon-number variance of the spectrally filtered
solitonic field. The full line is the ideal QNLSE at $\xi=4$, the dotted line is
the ideal QNLSE at $\xi=8$, the dashed line is at $\xi=4$ with the Raman
interaction at $T=300 K$ with 0.1 dB/km
(0.0236 dB/$z_0$ for $z_0=370m$) losses and dot-dashed line is
the latter case in the normal dispersion regime.
\label{fig2}}
\end{figure}
\begin{figure}
\caption{Scaled intensity spectrum variance versus propagation distance for $\phi(0,\tau)=\mbox{sech}(\tau)$
initial input pulses using the ideal QNLSE. \label{fig3}}
\end{figure}
\begin{figure}
\caption{Photon-number variance versus propagation distance 
and spectral filter cut-off frequency for $N=1$  
at $T=300 K$,$\, t_0\!\!=\!\!1$ ps. \label{fig4}}
\end{figure}
\begin{figure}
\caption{Scaled intensity spectrum variance versus propagation distance for $N=1.1$,$\phi(0,\tau)=N\mbox{sech}(\tau)$,
initial input pulses using the ideal QNLSE. By using the color map of Fig.~3,
the min/max values are clamped.\label{fig5}}
\end{figure}

\begin{thebibliography}{99}
\bibitem{gawbs}
An effect at low frequencies is GAWBS noise \cite{shelbylb:85} which differs significantly depending
on the fiber used. 
For this reason no attempt is made to include GAWBS contributions explicitly 
although one of the advantages of the approach used 
here is that such details can be included easily for specific fibers.
\bibitem{ramanfilter}
An interesting point is that the Raman effect provides an effective filter mechanism on the
dynamics of the nonlinear system and the noise added is temperature-dependent colored noise.
The Raman gain extends out to 40 THz but the
main contribution extends only to about 20 THz. 
For very short pulses this may provide unexpected properties.
This effective filter mechanism is quite general 
and arises from tracing out an interacting field
whose coupling with the field of interest is strongly frequency dependent.
\bibitem{heuristic}
It is expected from the earlier work of Gordon~\cite{gordon} on the perturbed classical
NLSE that vacuum fluctuations in the input pulse will produce a modulation
of the photon-number variance due to the dispersive radiation (or excited continuum). 
This can be seen by writing the ensemble averaged 
intensity dependent nonlinearity using the truncated Fokker-Planck equation 
for the Wigner representation~\cite{wigner}.
A heuristic argument can be made that an initial coherent pulse (a Gaussian distribution)
produces an effective increase in the nonlinearity
due to the 1/2 photon per mode vacuum fluctuations which modulate the intensity and in turn the spectrum.
\bibitem{numerical}
The simulations for all figures were carried out 
using a time window of 80ps which
determines the frequency resolution to be 12.5GHz.
\end{thebibliography}
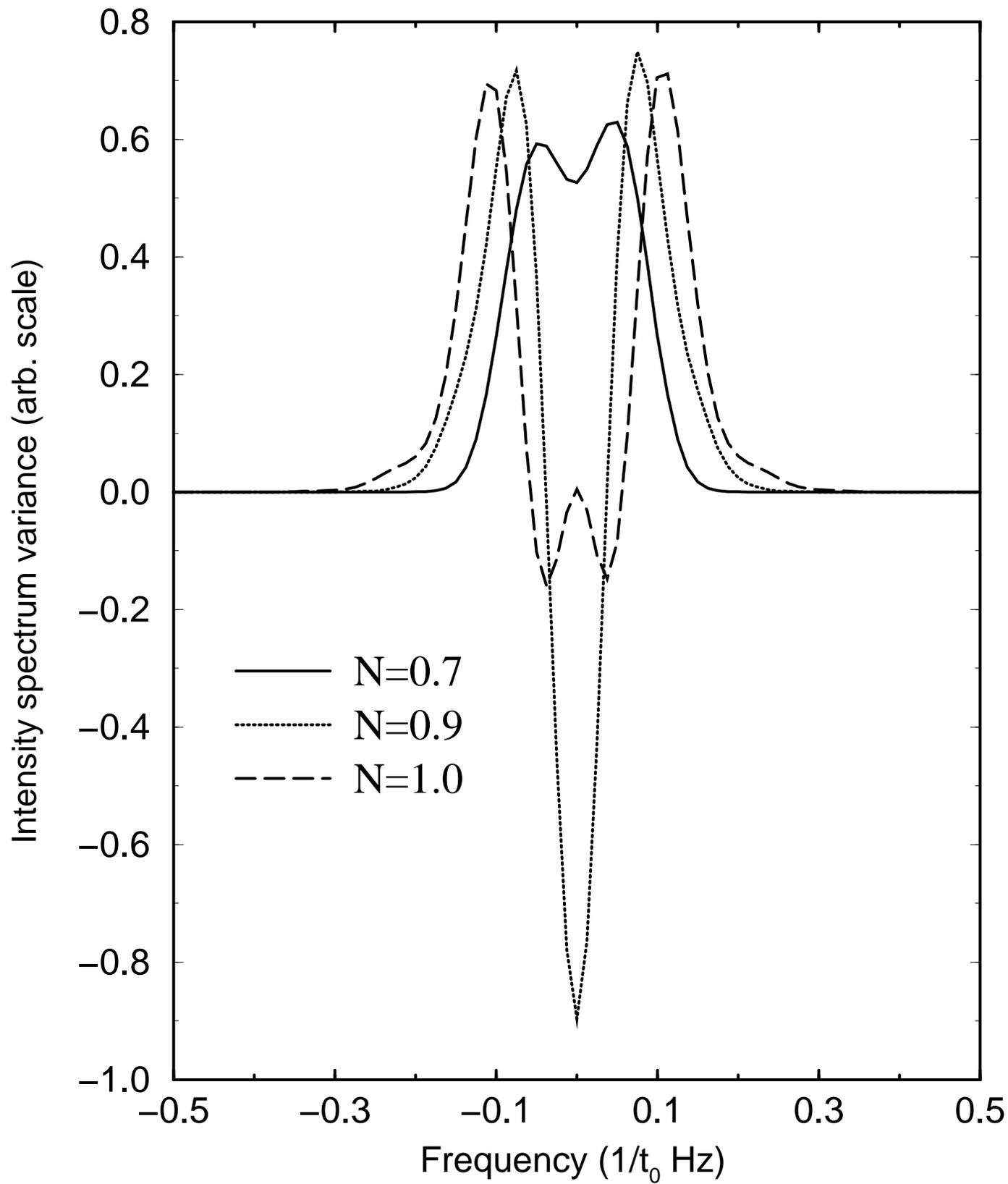
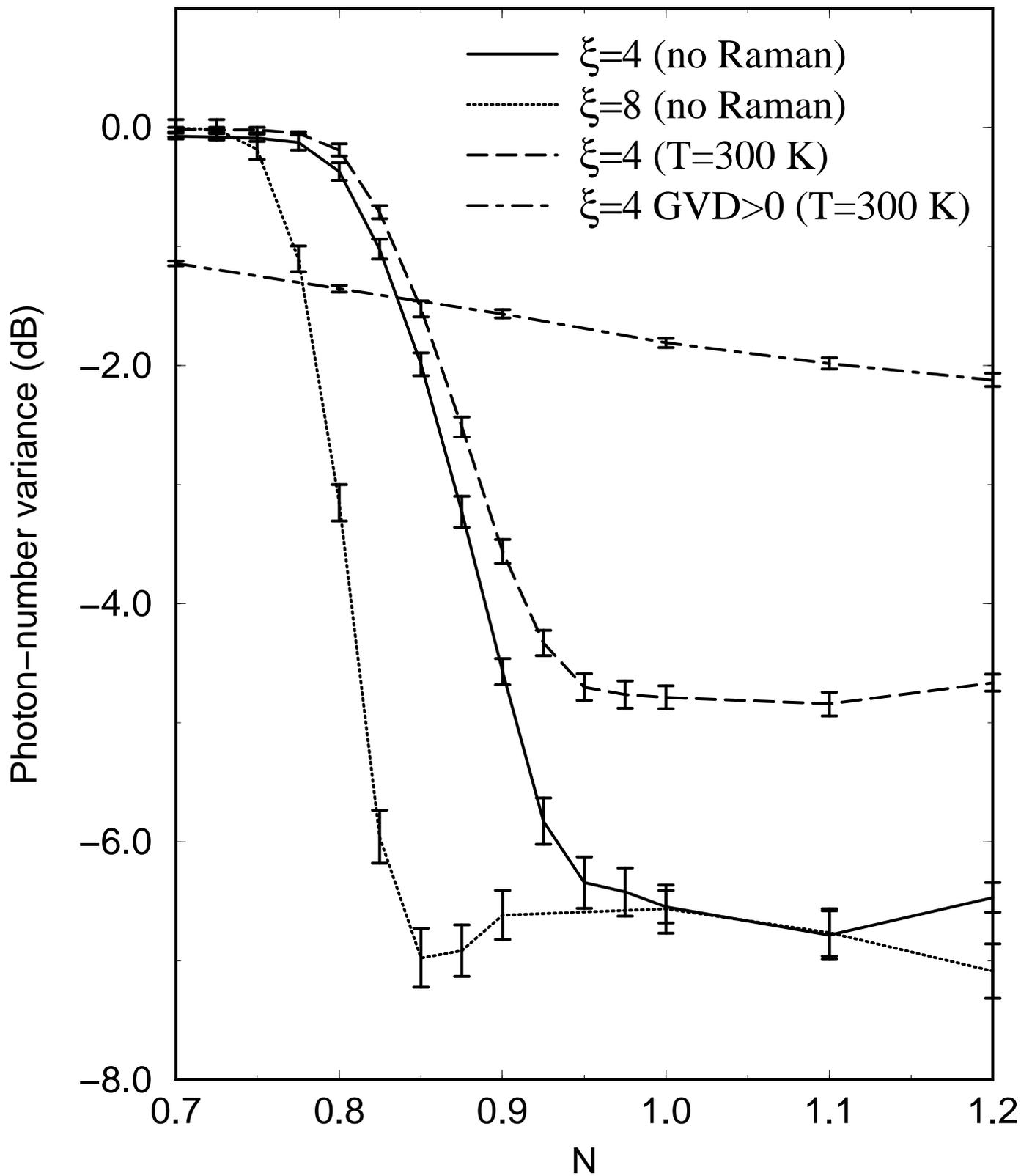
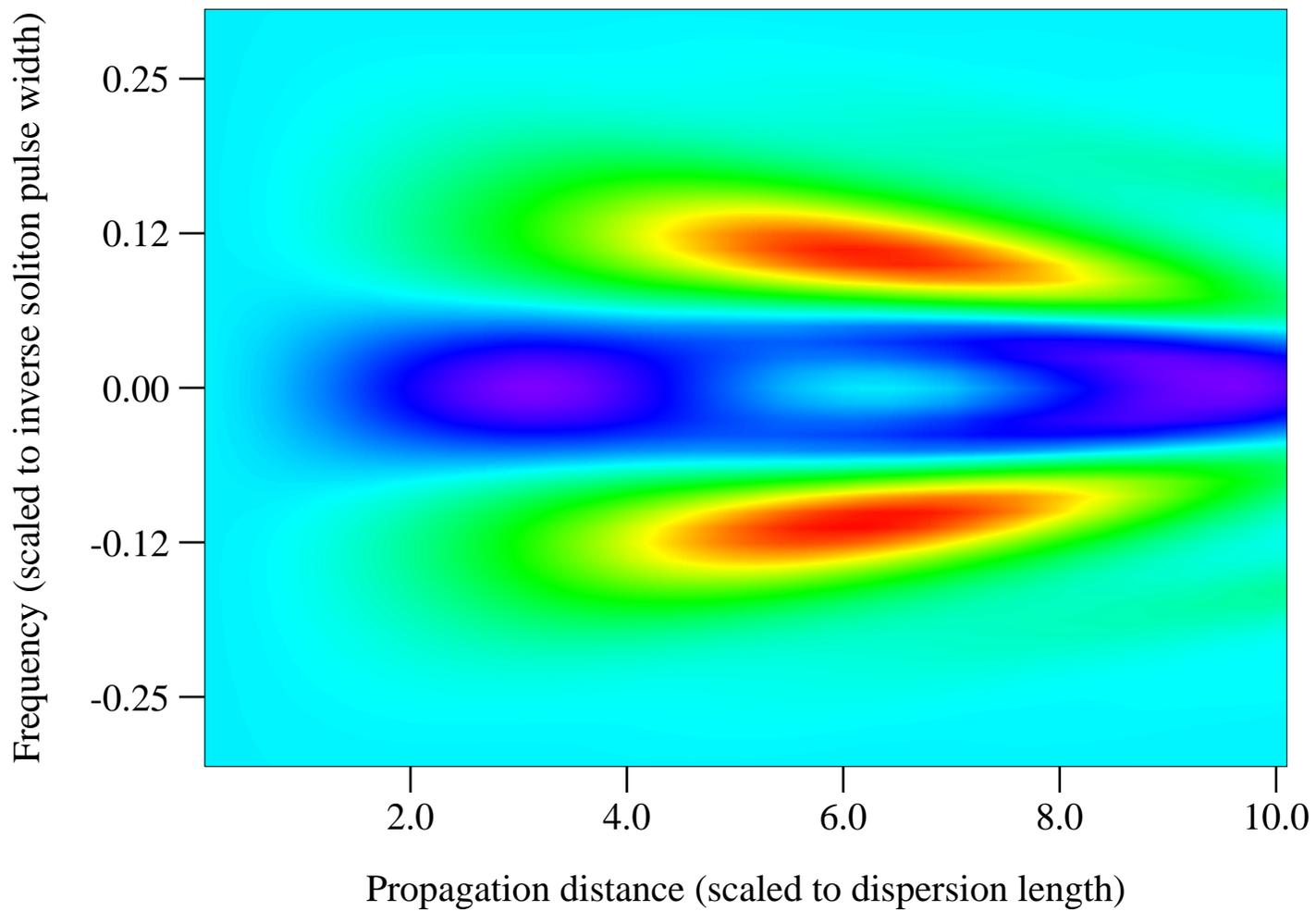
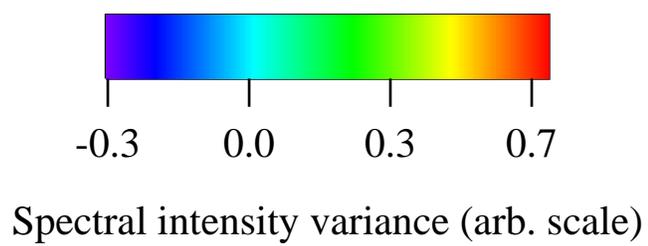
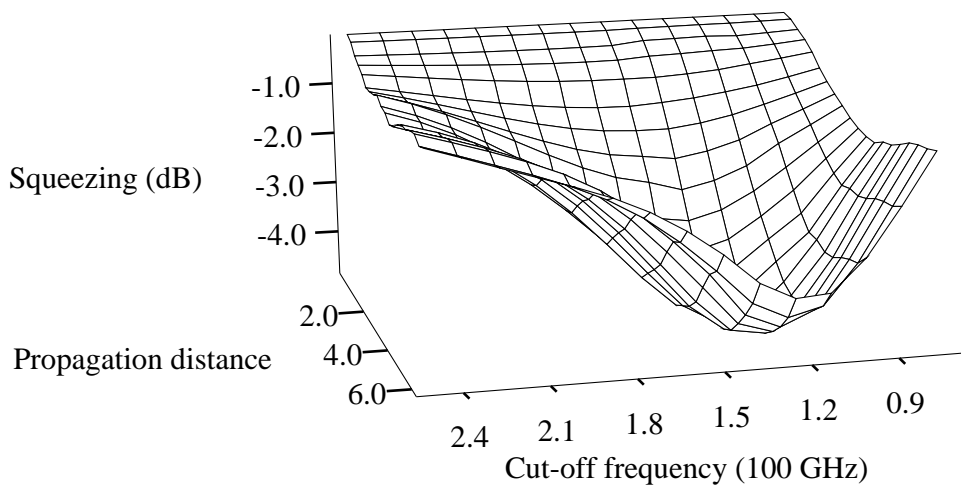
\end{document}